\documentclass[aps,prl,twocolumn,epsfig]{revtex4}
\usepackage{graphicx}

\begin{document}
\tighten

\def\tcn{T_c^{clean} }
\def\tcl{T_c(L)}
\def\tcd{T_{cond} }
\def\dt2{\delta T(L) }

\def\And{Andronishkashvili }
\def\ncri{non-classical rotational inertia }
\title{Quenched Dislocation Enhanced Supersolid Ordering}

\author{John Toner}
\address{Department of Physics and Institute of Theoretical Science,
University of Oregon, Eugene, OR 97403, USA}

\date{\today}

\begin{abstract}

I show using Landau theory that quenched dislocations can
facilitate the supersolid
(SS) to normal solid (NS) transition, making it possible for the transition
to occur  even if it does not in a dislocation-free crystal. I make detailed
predictions for the dependence of the SS to NS transition
temperature $T_c(L)$,
superfluid density 
and dislocation spacing $L$, all of which can be tested against
experiments. The
results should also be applicable to an enormous variety of other systems,
including, e.g., ferromagnets.
\end{abstract}


\maketitle

Recent
reports\cite{kim_nature_2004}  of
supersolidity - a crystal exhibiting  ``off-diagonal long-range order"
(ODLRO) 
\cite{leggett70, yang62}-in  solid
$^4$He raise many questions.
First, quantum Monte Carlo
simulations\cite{ceperley04}
find no supersolid phase.
Second,  the  temperature ($T$)
dependence\cite{kim_nature_2004} of the superfluid density
$\rho_S(T)$ in  the supersolid (SS)  differs from that in
the super{\it fluid} (SF), contradicting theory \cite{DGT}. Third,  no
specific heat anomaly  is seen at the SS to NS transition.

In this paper, I propose a resolution of these puzzles.  Since,
depending on the material,  {\it either} local compression {\it or} local
dilation increase the  local transition temperature
$T_c(\vec{r})$\cite{DGT}, and
since   edge dislocations  have regions of  {\it
both} types near their
cores\cite{landau_elasticity}, these defects induce, in {\it  all} 
materials, regions of  elevated $T_c$, as first noted   for
superconductors\cite{Nabitovskii}. ODLRO
therefore happens  at higher temperatures on the tangled network of
quenched dislocations  in
$^4$He crystals   than  in
the bulk, as in superconductors
\cite{Nabitovskii,Shantsev}, and
can occur even
if  the clean (dislocationless) lattice  remains normal down to $T = 0$.

Specifically,  the DGT\cite{DGT}
model with quenched dislocations  implies the following scenario: as
temperature $T$ decreases below what I'll call the  ``condensation''
temperature
$T_{cond}$, which  is {\it always} $>\tcn$, the transition temperature of the
clean (i.e., dislocationless) lattice, each  dislocation line in a tangled
network of them  nucleates  a cylindrical supersolid ``tube''  tangent
to it.
The radius of these tubes grows with decreasing
temperature.

We can think of places where
dislocations cross, making supersolid tubes overlap, as the ``sites''of a
random lattice. The sections of tube between these ``sites'' act as
ferromagnetic ``bonds''. The typical length of these bonds is
$L$, the mean dislocation spacing, which grows with annealing;
$L\rightarrow\infty$  for a clean crystal.
This random lattice does {\it not} develop
macroscopic supersolidity (or undergo {\it any} phase transition) at $\tcd$,
because the sites lack long-range phase coherence near $\tcd$. However,
as temperature is lowered further, such coherence {\it inevitably}
develops at  $T=\tcl$, with $\tcd >
\tcl  >
\tcn$.
Indeed, if  condensation occurs, long-range order
always develops (i.e.,
$\tcl>0$), {\it even if} the clean system {\it never orders!}
This ordering at $\tcl$  is the SS to NS   transition.

This picture is very similar to  Shevchenko's\cite{shev}.

\begin{figure}[b,h]
\vskip -0mm
\includegraphics[width=7cm]{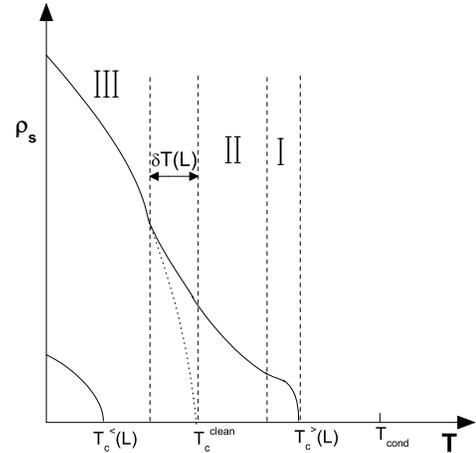}
\caption{The superfluid density versus temperature for a dislocated
solid in which the clean system {\it does} (top curve)  and {\it does not}
(bottom curve) have a transition. $\rho_s$ obeys eqns. (\ref{rhocrit}) and
(\ref{rhowell2}) in region (I) and (II) respectively, where (II) is 
defined by eqn. (\ref{window}). $\tcl$  for the cases $\tcn >$ and $< 0$
are, respectively, denoted in this figure by
$T^>_c(L)$ and $T^<_c(L)$, and given by eqns. (\ref{tcl}) and
(\ref{tcl2}) .
  }
\label{fig:1}
\end{figure}

Figure 1 plots the superfluid density $\rho_s(T)$. When $\tcn>0$, near
$\tcl$,
\begin{eqnarray}
\rho_S(T, L) = {A \over L^\chi} \left(1-{T \over T_c(L)}\right)^\nu  \quad
,
\label{rhocrit}
\end{eqnarray}
where $\nu\approx 2/3$ is the 3dXY correlation length exponent\cite{XY}, $\chi
=2({1-\nu \over 2-\nu})\approx {1 \over 2}$,
$A$ and
$T_0$ are independent of $L$,  $a$ is a lattice constant, and 

\begin{eqnarray}
T_c(L) =
T_c^{clean} + T_0 \left({a \over L}\right)^{1 \over
{2-\nu}}\quad , \quad &\tcn>0 ,
\label{tcl}
\end{eqnarray}

When  temperature $T$ is lowered into the range
\begin{eqnarray}
\delta T(L)\ll T-\tcn\ll\tcl-\tcn,
\label{window}
\end{eqnarray}
where
$
\delta T(L)\propto{1\over L}
$,
\begin{eqnarray}
\rho_S(T, L)  = A' {{(T - T_c^{clean})^{\nu-2} \over L^2} }\propto (T -
T_c^{clean})^{-{4 \over 3}}       ,
\label{rhowell2}
\end{eqnarray}
and $A'$ is an $L$-independent constant.
In the $L\rightarrow\infty$ limit, $\delta T(L) \propto {1\over L} \ll
\tcl - \tcn \propto L^{- {1\over {2-\nu}}} \approx L^{-{3 \over 4}}$ ,
ensuring a large window  of validity for  eqn. (\ref{rhowell2}).
Once $T < \tcn -\delta T(L)$, the tubes overlap,  the
entire volume becomes supersolid, and
$\rho_s$ is that of the clean system, completely independent of $L$, and so
obeys
\begin{eqnarray}
\rho_S(T)  \propto (
T_c^{clean}-T)^{{2 \over 3}}   .
\label{clean}
\end{eqnarray}

Note that the high temperature ($T > \tcn -\delta T(L)$) behavior
of $\rho_s(T,L)$
is strongly sample and annealing dependent (because
$L$-dependent), but the {\it
low}-temperature ($T < \tcn -\delta T(L)$) behavior is  sample and
annealing {\it independent}, and
identical to that of a clean sample.

Precisely such   behavior was recently  reported
\cite{chan2}. In
figure 2,    $\rho_s(T)$
data  from Chan's group\cite{thank} is plotted in the form $\rho_s^{-{3
\over 4}}$ versus $T$, which  Eqn. (\ref{rhowell2}) predicts should give a
straight line section, for $T$ satisfying eqn. (\ref{window}) .  The data
does indeed show such a straight section, although it  is fairly  short, 
and the error bars in this region  are large. 
More accurate measurements of $\rho_s(T)$ , and of the
dislocation spacing $L$ (by, e.g.,  ultrasonic
velocity and attenuation measurements
\cite{Beamish}) 
are clearly needed.
Alternatively, one could deduce the ratio of 
$L$'s in different samples by comparing the coefficients of
$(T_c^{clean}-T)^{-{4
\over 3}} $ in eqn. (\ref{rhowell2}), and using this ratio 
to test the predicted $L$ dependence of
$\rho_s(T, L)$ and $\tcl$ eqns. (\ref{rhocrit}) and(\ref{tcl}).

\begin{figure}[b,h]
\vskip -0mm
\includegraphics[width=7cm]{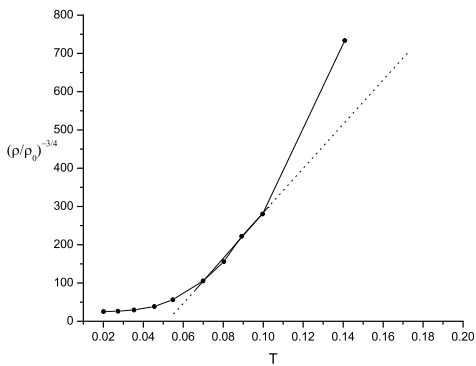}
\caption{Superfluid density versus temperature
data from Chan's
group\cite{thank}, plotted in the form $\rho_s^{-{3
\over 4}}$ versus $T$. The straight segment of this plot predicted by
Eqn. (\ref{rhowell2}) is indicated by the dashed line.
}
\label{fig:2}
\end{figure}

If the clean system does {\it not} order,
which I'll refer to as $\tcn < 0$,
$\tcl$ vanishes as $L
\rightarrow \infty$:

\begin{eqnarray}
T_c(L) = T_0 {a \over L}\quad , \quad \tcn<0 ,\
\label{tcl2}
\end{eqnarray}
where $T_0$ is another L-independent constant, a result
ref.\cite{shev}, could also be tested by
measurements and/or deductions (as described above) of
$L$. Eqn.(\ref{rhocrit}) still holds near $\tcl$, but now with
$\chi=2$.   Eqns (\ref{rhowell2}) and (\ref{clean}) never apply, since
$T=0$ intervenes above $\tcn$. The lower curve in
Figure 1 plots
$\rho_s(T)$ in this case.

The experimental situation is currently unclear.
The $\tcn <0$ scenario is supported by recent experiments\cite{Reppy},
showing   \ncri in unannealed $^4$He crystals, but none after
annealing. On the other hand, Chan's recent experiments\cite{chan2}, as
discussed above and in figure (2), suggest $\tcn > 0$. In these
experiments of ref. \cite{chan2}, single crystals still show
supersolidity, suggesting that dislocations, rather than grain
boundaries, are the responsible defects.

Also suggestive are  simulations\cite{Troyer}
which see supersolid order near screw dislocations.
Although screw dislocations 
do not, in the DGT model, couple to supersolid order, higher order terms
allow such coupling(\cite{JTfut}).

The absence of a specific heat anomaly in some experiments can be
explained  in this picture. For the case $\tcn < 0$, the
specific heat  near
$T_c(L)$ is given by:
\begin{eqnarray}
{C } \propto 
   {\left|{T \over T_c} - 1  \right|^{- \alpha}\over {L^{\left({4 - 3\nu
\over 2-\nu}\right)}}}\propto {\left|{T \over T_c} - 1  \right|^{-
\alpha}\over  L^{3 \over 2}}, 
 \tcn<0
\label{heat}
\end{eqnarray}
where $\alpha = -.0127$ is the specific heat exponent of the  3d $XY$ model
\cite{XY}. Clearly, the $\left|T - T_c  \right|^{- \alpha}$
singularity  vanishes as dislocation
density
$\rightarrow 0$ ($L \rightarrow \infty$), and so should be
seen only in dirty samples, not clean ones.

The ideas developed here are applicable to, e.g.,
ferromagnets\cite{Halperin} , which I'll treat elsewhere\cite{JTfut2}.

I'll now outline the derivation of these results.
My Hamiltonian is an isotropic\cite{iso}
version of that of
\cite{DGT}:
\begin{eqnarray}
H = \int d^3 r \left[{t(\vec{r}) \over 2} |\psi|^2 + {u \over
4}|\psi|^4 + {c \over 2 } |\vec\nabla\psi|^2 \right]
\label{H}
\end{eqnarray}
with
\begin{eqnarray}
t(\vec{r}) &=& t_0 + g u_{ii}(\vec{r})\  \quad .
\label{tr}
\end{eqnarray}
Here, $t_0(T)$ is a  decreasing function of temperature $T$
satisfying $t_0 (T_c^{clean}) = t_c^{clean} < 0$, where $t_c^{clean}$ is
the value of
$t_0$ at  the transition in the clean system, $u$
and $c$ are constants, and
$u_{ii}$ is the trace of the strain tensor.   

Thermal fluctuations in $u_{ii}$  have no
effect on the critical properties of the superfluid density and specific
heat at the transition\cite{DGT};
I will henceforth ignore them, and
focus only on strains due to quenched dislocations.

The clean
model will {\it not} have a
$NS
\rightarrow SS$ transition if $T_c^{clean} < 0$.  When $T_c^{clean} > 0$,
I'll assume (as usual) that
$t_o(T) = \Gamma {\left( T - T_c^{clean}\right) \over \left| T_c^{clean}
\right| }$, where $\Gamma$ is a constant,  near $T = T_c^{clean}$.  For a
straight edge dislocation running along the
$z$-axis with
Burgers vector
$\vec{a} $ along the $y$-axis,
$
u_{ii} = {4\mu \over 2 \mu + \lambda} {ax\over r_{\perp}^2} = {4\mu \over 2
\mu + \lambda}
{a \cos \theta \over r_{\perp}}$ \cite{landau_elasticity},
where $\mu$ and $\lambda$ are the Lame elastic constants\cite{iso}.
Inserting this into eqn. (\ref{tr}) gives
\begin{eqnarray}
t(\vec{r}) &=& t_0 + {g^\prime \cos \theta \over r_{\perp}}\  \quad ,
\label{trdipole}
\end{eqnarray}
where  $g^\prime\equiv ga \left( 4\mu \over 2\lambda +
\mu\right)$\cite{iso}.

Naively, the system is supersolid in those regions where $t({\vec r})<0$.
Actually, the mean field transition occurs  when the minimum
energy
$\psi
\left(\vec{r}\right)$ first becomes non-zero.  The temperature
at which this  occurs is
$T_{cond}$.

The Euler-LaGrange equation for eqn. (\ref{H}) is
\begin{eqnarray}
\nabla^2{\psi} = {t (\vec{r}) \over c} \psi + {u \over c} \psi^3  \quad .
\label{elg}
\end{eqnarray}
As noted in \cite{Nabitovskii},this
equation first has  non-trivial ($\psi\neq0$)
solutions when $t_0$ drops below a critical value
$
t_{cond} \equiv -{2mE_0 \over \hbar ^2},
$ where $E_0$ is the quantum mechanical ground state energy of a   particle of
mass
$m$ moving  in the 2d dipole potential
$
V (\vec{r}) =  - {p \cos \theta  \over 2m r_{\perp}}
$ with
$
p \equiv  {\hbar^2 g^{\prime} \over 2mc}
$.
Variational treatments \cite{Nabitovskii,Griffiths}, show  that
$ E_0 = - \gamma {mp^2 \over \hbar ^2}
$ where
$0.24 < \gamma < 2$.
So a single dislocation line will, in mean field theory,
  order  once
$t_0 < t_{cond} =  { \gamma g \prime ^2 \over 2c}.
$ Using $t_0 = \Gamma \left({T - T_c^{clean} \over  \left|
T_c^{clean}\right|  }\right)$, this implies
$
T_{cond}  =   T_c^{clean} +  { \gamma g \prime ^2 \over 2 \Gamma c} \left|
T_c^{clean}\right|
>  T_c^{clean}$, and
$\tcd>0$ ,   even if $ T_c^{clean} < 0$, if ${\gamma
g^{\prime 2} \over 2\Gamma c} >1 $.
Hence, condensation onto  dislocations  {\it can} happen,  {\it
even} when the clean system does not order.

However, a
one-dimensional system like a single dislocation line cannot order.
To order, these 1d ``tubes''  must cross-link into a
three-dimensional network.  The typical tube
length is $L$, the inter-dislocation distance.

On  length scales $\gg$ the tube radius $a_c (t)$, but $\stackrel
{<}{\sim} L$, the only important variable is  ``Goldstone mode''; i.e.,
the phase
$\theta (\vec{r})$ of  $\psi(\vec{r}) \equiv
\left| \psi(\vec{r})\right|e^{i\theta (\vec{r})}$. In the tube  between
crosslink sites $i$ and $j$, $\theta$, on these long length scales,
depends only on distance $s$ along the tube. This leads to a $1d$ 
Hamiltonian for this tube:
\begin{eqnarray}
H_{1d}(\{\theta(s)\}) = K_{1d}(T)\int^L_0   \left(\partial _s \theta \right)^2
ds  \quad .
\label{H1d}
\end{eqnarray}

 From this, I can obtain an effective Hamiltonian $H_{eff} (\theta_i ,
\theta_j)$ coupling the $\theta$'s on sites $i$ and $j$ by integrating out the
$\theta$'s along the tube:
\begin{eqnarray}
e^{-\beta H_{eff} (\theta_i , \theta_j)} = \sum^{\infty}_{n = - \infty}
\int_n D
\theta(s)  e^{-\beta H_{1d} (\{\theta(s)\})}
\label{He1}
\end{eqnarray}
where the functional integral $\int_n D\theta (s)$ on the right hand side is
taken with $\theta (s)$ satisfying the boundary conditions
$
\theta (0) =\theta_i$ , $\theta(L) = \theta_j + 2 \pi n$, where the
summation integer $n$  in eqn.
(\ref{He1}) reflects the $2\pi$ periodicity in $\theta$.
%

Each of the functional integrals $\int_n  D \theta(s)$ in eqn. (\ref{He1})
can most easily be done by
rewriting $\theta(s)$ as follows:
\begin{eqnarray}
\theta (s) = \theta_i + \left({\theta_j - \theta_i +2\pi n\over L} \right) s +
\delta \theta(s) ,
\label{theta_min}
\end{eqnarray}
where the new  integration variable $\delta \theta (s)$ satisfies
the boundary conditions $
\delta \theta (0) = \delta \theta (L) = 0 .
$  This gives
\begin{eqnarray}
e^{-\beta H_{eff} (\theta_i,\theta_j)} &=& \sum^{\infty}_{n = - \infty}
e^{-\beta {K_{1d}\over L}(\theta_i - \theta_j + 2\pi n)^2} \nonumber\\ &&
\times
\int D
\delta \theta(s) e^{-{\beta K_{1d}}\int^L_0 ds \left(\partial_s \delta
\theta(s)\right)^2}  .
\label{He2}
\end{eqnarray}
  The $\int D\delta \theta$ in eqn. (\ref{He2}) is independent of
$\theta _i$ , $\theta _j$ and $n$ (since  the boundary conditions
on $\delta \theta$ are), and so is only an overall
multiplicative constant in $e^{-B H_{eff}}$, which only adds an
irrelevant  constant $C $ to
$H_{eff} (\theta _i, \theta _j)$.  Hence,
$H_{eff} (\theta _i,
\theta _j)$ becomes a ``periodic Gaussian''
\cite{Villain}
\begin{eqnarray}
H_{eff} (\theta _i, \theta _j) = V_v (\theta _i-\theta _j; J)\nonumber\\  =
-k_BT ln \left(\sum^{\infty}_{n = - \infty}e^{- {J \over k_BT}(\theta _i-
\theta _j+ 2 \pi n)^2} \right) +C  \quad ,
\label{Villain}
\end{eqnarray}
with the ``Villain'' coupling
\begin{eqnarray}
J \equiv {K_{1d} \over L}  \quad .
\label{J1}
\end{eqnarray}

Adding up $H_{eff} (\theta _i, \theta _j)$ for {\it all} of the bonds gives
a  model for all of the ``sites'' (cross links of  tubes  ):
\begin{eqnarray}
H_{eff} (\{\theta_i\})=  \sum_{bonds}V_v (\theta _i-\theta _j; J)  \quad .
\label{V2}
\end{eqnarray}

Although these couplings $J$
will be random, due to
the  random bond lengths of the tubes,
   such ``random $T_c$'' disorder is irrelevant in the
RG\cite{Harris}, and
can be ignored.

This Villain model (\ref{V2}) orders  at a
temperature
  $T_c = O(J/k_B)$; I will now use this to
determine $T_c(L)$.

Consider first $T_c^{clean}< 0$. In this case, provided $T_{cond}\geq
0$, so that $K_{1d}(T)\neq 0$, we can,
for  $L\rightarrow \infty$, estimate
$T_c$ by replacing $K_{1d}(T)$  in eqn. (\ref{J1})  with its finite, non-zero,
$T = 0$  value  $K_{1d}(T= 0)\equiv K_0$.  This gives eqn. (\ref{tcl2}) with
$
T_0= {K_0 \over k_B} .
$  Note  that taking
$K_{1d}(T)
  \rightarrow K_0$ in eqn. (\ref{J1}) is valid
since
$T_c (L \rightarrow \infty)\rightarrow 0$.

For the case $T_c^{clean}> 0$, the radii $a_c(T)$ of the tubes of
supersolid  diverge as
$T\rightarrow T_c^{clean}$.  To see this, note that the locus on
which $t(\vec{r}_{\perp})$  eqn. (\ref{trdipole})  is equal to $t_c$
is  $ \cos \theta = {r (t_c - t_0 ) \over g\prime} $ which, for $t_0 >
t_c$ and $p >(<) $ $0$  is a circle passing through the origin,
centered on the   negative (positive) $x$-axis of radius
\begin{eqnarray}
a_c(T) = {g\prime  \over 2 (t_0 - t_c)}  \quad .
\label{radius}
\end{eqnarray}
Inside this circle, $t(\vec{r}_{\perp}) < 0$, so, naively, this
boundary (\ref{radius}) defines the  supersolid tube. As $T
\rightarrow T_c^{clean}$ from above, $t_0 \rightarrow t_c$ and so
$a_c(T)$  diverges: $a_c(T) \propto {1 \over T - T_c^{clean}}$. Of
course, this argument  ignores  the $\nabla ^2 \psi$ term in eqn.
(\ref{elg}).
  However, since $a_c(T)
\rightarrow \infty$ as $T
\rightarrow T_c^{clean}$,  $\psi$ varies slowly in space, and  we can neglect
the $\nabla^2 \psi$ term in
eqn. (\ref{elg}) and simply balance the other
two terms.

We can include fluctuations  in this ``local equilibrium'' approximation
simply by replacing the local superfluid density
$\rho _s (\vec{r})$ by its value  in a {\it uniform} system whose
value of
$t$ equals the local $t(\vec{r})$,  provided $a_c(T)>> \xi (T)\propto (T
- T^{clean}_c)^{-\nu}$, where $
\xi (T)$ and $\nu \approx {2 \over 3}$ are  the  correlation length and
its critical exponent in the
the clean system.
    Since $\nu < 1$,
$
a_c(T)$  eqn.  (\ref{radius}) is indeed $ >> \xi (t)
$ as $T \rightarrow T^{clean}_c$ from above.
This implies that the local superfluid density $\rho_s (\vec{r})$
for $T$ near, but slightly above, $T^{clean}_c$, is
\begin{eqnarray}
\rho^{local}_s (\vec{r}) = B (t_c - t(\vec{r}))^\nu \quad ,
\label{rho1}
\end{eqnarray}
where $B$ is a constant, and I've used  the Josephson relation $\rho_s
\propto \xi^{-1}$\cite{Josephson}.
This $\rho_s$ acts as the $3-d$  ``spin-wave
stiffness'' for the phase
$\theta(\vec{r})$; that is,
\begin{eqnarray}
H_{3d} = {1 \over 2} \int d^3 r K_{local}(\vec{r})\left| \vec{\nabla}
\theta
\right|^2 \quad ,
\label{Kdef}
\end{eqnarray}
with
$
K_{local}(\vec{r}) =  {\hbar ^2 \over  m^2} \rho^{local}_s (\vec{r})
$. In the case of a straight edge dislocation,  taking
$t(\vec{r})$ from eqn. (\ref{trdipole}), $\rho_s (\vec{r}) $  by
eqn. (\ref{rho1}) and $\theta(\vec{r})$  to
vary only with distance $s$ along the dislocation line, the 1d spin wave
stiffness $K_{1d}$ becomes:
\begin{eqnarray}
K_{1d} &=&  {\hbar^2 \over m^2} \int d^2 r_{\perp} \rho^{local}_s
(\vec{r}_{\perp})
\quad .\label{K1d}
\end{eqnarray}

Since $t(\vec{r}_{\perp})$ is constant on circles of fixed radius
$a$, passing through the origin, with their centers on the $x$-axis,
and is given by: $ t({\vec r}_\perp) =  {p \over 2a} -  t_0   , $
I'll change variables of integration in eqn. (\ref{K1d}) to $a$. The
area of the interval $[a ,  a + da]$ is  the difference $2 \pi ada$
between the areas of the corresponding circles, so I can rewrite
eqn. (\ref{K1d}) as
\begin{eqnarray}
K_{1d}(T) &=&{\frac{\pi B\hbar ^{2}}{m^{2}}}\int_{0}^{\frac{p}{2t_{0}}%
}\left( {\frac{p}{2a}}-\delta t_{0}\right) ^{\nu }ada \nonumber\\
&=&C(\frac{\mu }{\lambda })B\delta t_{0}^{\nu -2}g^{2}a^{2}{\frac{\hbar ^{2}%
}{m^{2}}} \quad , 
\label{K1d2}
\end{eqnarray}
where  $\delta t_0 \equiv t_0 - t_c$, and
$C (x) \equiv   \left({3.184 x \over 2x + 1}\right)^2$ for
$\nu = 2/3$.

Since $\delta t_0 \propto T - T^{clean}_c$, eqn. (\ref{K1d2}) implies that
$K_{1d}(T)
\propto (T - T^{clean}_c)^{\nu - 2}$.  Using this
$K_{1d}(T)$ in my earlier expression (\ref{J1}) for $J$,  and then equating
the result to $k_B T$, gives  eqn. (\ref{tcl}) for $\tcl$.

As  $T$ drops further,  eventually
$J(T, L)$ will be $ \gg k_B T$. This is guaranteed to happen, since $T$
can get within roughly $\dt2\propto {1 \over L}$
of $\tcn$ before eqn. (\ref{K1d2}) breaks down. Since $K_{1d}(\tcn +\dt2)
\propto L^{2-\nu}$;  $J(\tcn +\dt2)= {K_{1d}\over L} \propto
L^{1-\nu}\rightarrow\infty$ as $L\rightarrow\infty$, since $\nu < 1$.
In
this limit, the phase order on the ``sites'' of the dislocation network
is nearly perfect, and the standard relationship between the
{\it macroscopic} (as opposed to the local) $\rho_s$ and 3d spin wave
stiffness  implies:
\begin{eqnarray}
\rho_S(T, L) = J(T, L)/L \times O({m^2 \over \hbar^2})
   \label{rhowell1}
\end{eqnarray}
which,  using
(\ref{J1}) for
$J(T,L)$, implies eqn. (\ref{rhowell2}).
Standard results for the
model (\ref{Villain}), (\ref{V2}) with the $T$ and $L$-dependent $J$
found above then gives the behaviors of
$T_c(C)$,$\rho _s (T)$, and the specific heat
$C(T)$ quoted earlier.

I thank A. Dorsey, D. Belitz, Q. Li, D. Thouless,
G. Dash, J. Beamish, P. Goldbart, and D. Griffiths for
valuable discussions, and
the KITP, Santa Barbara, CA; the
Aspen Center for Physics, Aspen, CO; and   the Marsh
Cottage Institute, Inverness, CA, for their hospitality.


\end{document}